\documentclass[prl,showpacs,floatfix,twocolumn]{revtex4}
\usepackage{graphicx}
\usepackage{dcolumn}

\begin{document}


\title{Strong-coupling superconductivity in
nickel-based LaO$_{1-x}$F$_x$NiAs}

\author{Z. Li$^1$, G. F. Chen$^1$, J. Dong$^1$, G. Li$^1$, W. Z. Hu$^1$,
D. Wu$^1$, S. K. Su$^1$, P. Zheng$^1$, T. Xiang$^{1,2}$, N. L.
Wang$^1$}

\author{J. L. Luo$^1$}
\email{jlluo@aphy.iphy.ac.cn}

\affiliation{$^1$Beijing National Laboratory for Condensed Matter
Physics, Institute of Physics, Chinese Academy of Sciences, Beijing
100190, China}

\affiliation{$^2$Institute of Theoretical Physics, Chinese Academy
of Sciences, P.O.Box 2735, Beijing 100190, China}

\date{\today}

\begin{abstract}

A series of layered nickel-based LaO$_{1-x}$F$_x$NiAs compounds with
x=0 to 0.15 are synthesized by solid state reactions. The pure
LaONiAs exhibits bulk superconductivity with Tc $\sim$ 2.75 K.
Partial substitution of oxygen by fluorine increases the transition
temperature to $\sim 3.8$ K. The LaO$_{0.9}$F$_{0.1}$NiAs sample
shows a sharp superconducting transition and a sharp specific heat
jump at the critical temperature. The magnitude of the specific heat
jump is much larger than that expected from the weak-coupling BCS
theory, indicating that this superconductor is in the strong
coupling regime. Furthermore, the temperature dependence of the
specific heat deviates strongly from the theoretical result for the
single-band s- or d-wave superconductor, but shows some character of
a multi-gap system.

\end{abstract}



\maketitle

Layered transition metal oxypnictides LaOMPn (M=Mn, Fe, Co, and
Ni, Pn=P and As) have attracted great attention recently due to
the discovery of superconductivity in the Fe- and Ni-based
systems. The superconductivity was first reported in Fe-based
LaOFeP with transition temperature T$_c\sim$4 K which increases to
7 K with F$^-$ doping,\cite{Kamihara06} later in Ni-based LaONiP
with T$_c\sim$3 K.\cite{Watanabe} With the replacement of P by As
and partial substitution of O by F in the Fe based compound to
yield LaO$_{1-x}$F$_x$FeAs, T$_c$ could raise to 26 K.
\cite{Kamihara08} At present much efforts have been devoted to the
Fe-based systems.\cite{Lebegue,Singh,fang,Haule,Ma,chen,wen}. It
is found that the undoped compound LaOFeAs itself is not
superconducting, but undergoes a spin-density-wave (SDW)
transition at 150 K. Upon fluorine doping, the SDW instability is
suppressed, and meanwhile the superconductivity starts to
appear.\cite{Dong} It is of great interest to see if similar
phenomenon could appear in the Ni-based superconducting compound
LaONiP when P is replaced by As and O is partially substituted by
F.

In this work, we present an extensive investigation on the
physical properties of high quality LaO$_{1-x}$F$_x$NiAs ($x=0\sim
0.15$) superconductors. Unlike LaOFeAs, we found that the pure
LaONiAs exhibits bulk superconductivity with Tc $\sim$ 2.75 K.
Partial substitution of oxygen by fluorine only increases the
transition temperature slightly, however it dramatically improve
the superconducting quality. The superconducting transition
temperature becomes extremely narrow with the width only about
0.05K when the F content is higher than 0.06, and the
superconducting volume fraction is extremely high. The very high
quality of the sample enables us to precisely determine the
superconducting parameters. Very sharp specific heat jumps at
superconducting transition temperatures were observed, and
detailed analysis of the specific heat data indicates that
LaO$_{1-x}$F$_x$NiAs is a strong-coupling and multi-gap
superconductor.

\begin{figure}
\scalebox{0.7}{\includegraphics{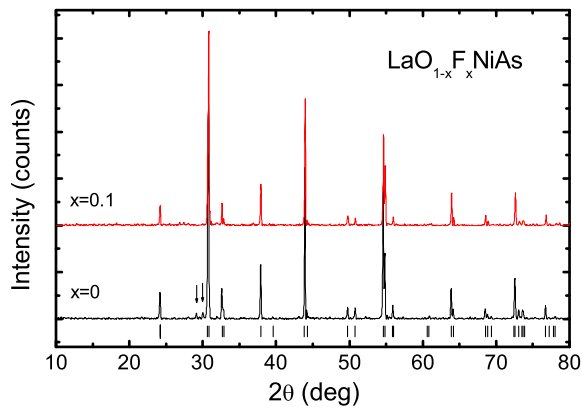}}
\caption{\label{fig:XRD} (Color online) The X-ray powder
diffraction patterns of LaONiAs and LaO$_{0.9}$F$_{0.1}$NiAs. The
bars at the bottom show the calculated Bragg diffraction positions
of LaONiAs. Two tiny peaks marked by arrows are from impurity
phase for LaONiAs. No impurity phase is detected for
LaO$_{0.9}$F$_{0.1}$NiAs. }
\end{figure}

The samples were synthesized by the solid state reaction using NiO,
Ni, As, La, and LaF$_{3}$ as starting materials. LaAs was prepared
by reacting La chips and As pieces at 500 $^{\circ}C$ for 15 hours
and then 850 $^{\circ}C$ for 2 hours. The raw materials were
thoroughly grounded and pressed into pellets. The pellets were
wrapped into Ta foil and sealed in an evacuate quartz tube under
argon atmosphere. Is was then annealed at 1150 $^{\circ}C$ for 50
hours. The resulting samples were characterized by the powder X-ray
diffraction (XRD) with Cu K$\alpha$ radiation at room temperature.
Figure \ref{fig:XRD} shows the XRD patterns for LaONiAs and
LaO$_{0.9}$F$_{0.1}$NiAs. It is found the XRD patterns are well
indexed on the basis of tetragonal ZrCuSiAs-type structure with the
space grounp P4/nmm. Only two extra tiny peaks were detected for
LaONiAs but no obvious impurity phase was detected for
LaO$_{0.9}$F$_{0.1}$NiAs. The lattice parameters are $a$=0.4119 nm,
$c$=0.8180 nm for LaONiAs, and $a$=0.4115 nm, $c$=0.8169 nm for
LaO$_{0.9}$F$_{0.1}$NiAs, respectively.

The electrical resistivity was measured by the standard 4-probe
method. The ac magnetic susceptibility was measured with a
modulation field in the amplitude of 10 Oe and a frequency of 333
Hz. The Hall coefficient measurement was done using a five-probe
technique. The specific heat measurement was carried out using a
thermal relaxation calorimeter. The field dependence of the
thermometer and heat capacity of addenda were carefully calibrated
before measurement. All these measurements were preformed down to
1.8K in a Physical Property Measurement System(PPMS) of Quantum
Design company.

\begin{figure}
\scalebox{0.65}{\includegraphics{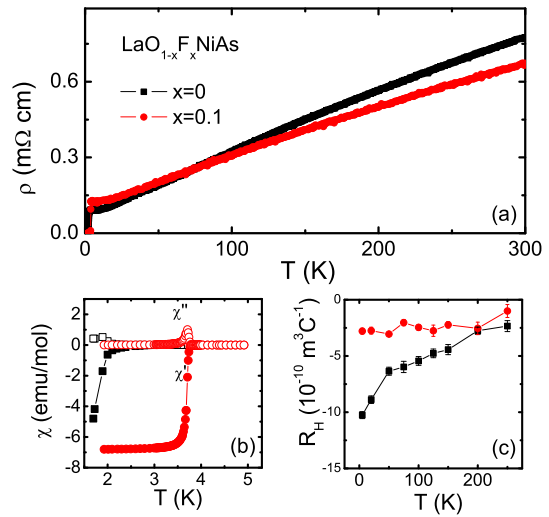}} \caption{\label{fig:2}
(Color online) (a)Temperature dependence of the resistivity $\rho$
for LaONiAs and LaO$_{0.9}$F$_{0.1}$NiAs, respectively. (b)The
real and the imaginary parts of ac susceptibility below 5K for
LaONiAs (open and closed squares) and LaO$_{0.9}$F$_{0.1}$NiAs
(open and closed circles), respectively. (c) Hall coefficient
versus temperature for the two samples.}
\end{figure}

Figure \ref{fig:2}(a) shows the temperature dependence of the
resistivity $\rho$ for LaONiAs and LaO$_{0.9}$F$_{0.1}$NiAs from
1.8 to 300 K at zero field. The normal state resistivity for both
samples is metallic, different from that observed in LaOFeAs. In
LaOFeAs, the resistivity exhibits a SDW transition at $\sim$ 150K,
a minimum at $\sim$ 100K, and an upturn in low
temperatures\cite{Kamihara08, Dong}. The onset superconducting
transition occurs at 2.75 K for LaONiAs and 3.8 K for
LaO$_{0.9}$F$_{0.1}$NiAs, respectively. The transition temperature
of LaONiAs is close to that of LaONiP with Tc$\sim 3 K$, but
differs from LaOFeAs which is non-superconducting with a SDW
transition at $\sim$ 150K. The superconducting transition width
for LaO$_{0.9}$F$_{0.1}$NiAs is $\sim$ 0.05 K, much narrower than
any other superconductor in this family, indicating a high
homogeneity of the superconducting phase.

The bulk superconductivity in these samples can be confirmed by
magnetic susceptibility measurements. Figure \ref{fig:2}(b) shows
the real ($\chi'$) and imaginary ($\chi"$) ac susceptibility
around T$_c$. Both LaONiAs and LaO$_{0.9}$F$_{0.1}$NiAs samples
become diamagnetic below $T_c$. The transition of
LaO$_{0.9}$F$_{0.1}$NiAs is much steeper than that of LaONiAs. For
pure LaONiAs, $\chi'$ begins to drop below 1.8 K, similar to
LaONiP \cite{Watanabe}. However, for LaO$_{0.9}$F$_{0.1}$NiAs,
$\chi'$ is already saturated below 3K. The absolute value of the
diamagnetic susceptibility is about 3 times larger than that
reported on LaO$_{1-x}$F$_{x}$FeAs, LaOFeP, and LaONiP samples. It
indicates that the volume fraction of superconducting phase in
this sample is very high.

The temperature dependence of the Hall coefficient R$_H$ for
LaO$_{1-x}$F$_x$NiAs with x=0 and 0.1 is shown in the inset of Fig.
\ref{fig:2}(c). The negative R$_H$ indicates that the charge
carriers are dominantly electron-type, same as in
LaO$_{1-x}$F$_x$FeAs\cite{chen}. The absolute value of R$_H$ for
LaO$_{0.9}$F$_{0.1}$NiAs is more than one order of magnitude smaller
than that of LaO$_{0.9}$F$_{0.1}$FeAs, indicating that
LaO$_{1-x}$F$_x$NiAs system has relatively higher carrier density.
This can be explained by the fact that Ni$^{2+}$ (3d$^8$)
contributes two more electrons than LaO$_{0.9}$F$_{0.1}$FeAs. The
Fermi energy shifts up comparing with that of Fe$^{2+}$ (3d$^6$) in
LaO$_{1-x}$F$_x$FeAs, the hole bands tend to be fully filled, and
the electron bands dominate the conductivity\cite{fang}.

\begin{figure}
\scalebox{0.60}{\includegraphics{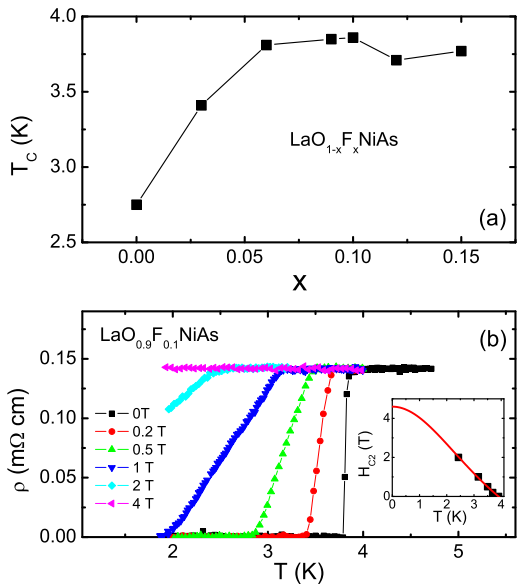}} \caption{\label{fig:3}
(Color online) (a) Doping dependence of the superconducting
transition temperature T$_c$ for LaO$_{1-x}$F$_x$NiAs. (b)
Temperature dependence of the resistivity of
LaO$_{0.9}$F$_{0.1}$NiAs at different fields. The inset shows the
temperature dependence of the upper critical magnetic field.}
\end{figure}

Figure \ref{fig:3}(a) shows T$_c$ as a function of x for
LaO$_{1-x}$F$_x$NiAs. With increasing x, T$_c$ first increases from
2.75 for x=0 to 3.78K for x = 0.06, and then remains almost constant
up to x=0.15. This doping dependence of T$_c$(x) is similar to that
of the LaO$_{1-x}$F$_x$FeAs system\cite{Kamihara08}. It suggests
that the two systems may have the same superconducting mechanism.
The superconducting transition width is about 0.05K when x is higher
than 0.06.

Figure \ref{fig:3}(b) shows the field dependence of the
resistivity of LaO$_{0.9}$F$_{0.1}$NiAs up to 4T. The transition
temperature Tc shifts to lower temperature in fields. The
transition width is gradually broadened, similar as for LaOFeAs
\cite{chen}. The field-induced broadening may result from the
large anisotropy of the upper critical field H$_{c2}$ for H//c and
H//ab plane. A 4T magnetic field suppresses T$_c$ down below 1.8K.
Using the onset superconducting transition temperature, the zero
temperature upper critical field H$_{c2}$(0) can be estimated with
the formula ${H_{c2}(T)}=H_{c2}(0)(1-t^2)/(1+t^2)$ \cite{chen,
wen}, where t is the reduced temperature t=T/T$_c$. By fitting, we
find that H$_{c2}(0) \sim$ 4.6 T [see inset of Figure 3(c)], which
is about 10 times smaller than the corresponding value for 10$\%$
F-doped LaOFeAs (H$_{c2} \sim$ 54 T).\cite{chen} The relatively
lower H$_{c2}$ in LaO$_{0.9}$F$_{0.1}$NiAs is due to a lower T$_c$
in this material.

Specific heat(C) measurement is a powerful tool to detect the bulk
properties of a superconductor both in the superconducting and
normal states. Generally, the specific heat is dominated by the
phonon contribution and it is difficult to separate electronic
contribution from the total specific heat in high temperatures.
Therefore, there is a large uncertainty in the determination of the
characteristic parameters, such as the normal state electronic
specific coefficient $\gamma_n$ and the specific heat jump at T$_c$,
if the measurement of the specific heat is done for a superconductor
with high $T_c$. For example, in LaO$_{0.9}$F$_{0.1}$FeAs with Tc
$\sim 20 K$, the specific heat anomaly at T$_c$ has not been
observed at zero field.\cite{wen}. However, the high superconducting
volume fraction and the low transition temperature in our
LaO$_{0.9}$F$_{0.1}$NiAs sample provides a good opportunity to
determine these superconducting characteristic parameters
accurately.

\begin{figure}
\scalebox{0.60}{\includegraphics{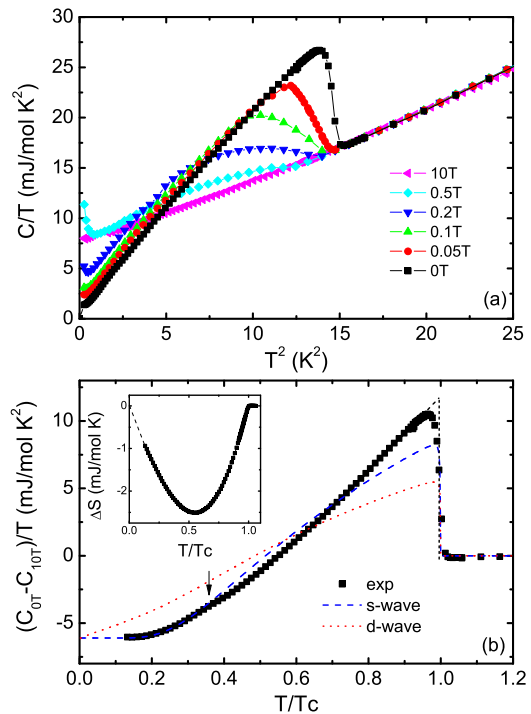}}
\caption{\label{fig:HC} (Color online) (a) C/T versus T$^2$ of
LaO$_{0.9}$F$_{0.1}$NiAs under selected magnetic fields. (b)
Temperature dependence of the difference in the specific heat
coefficient between zero field and 10T, (C$_{0T}$-C$_{10T}$)/T.
The dash and dot lines are theoretical curves for the BCS s- and
d-wave superconductors respectively. Inset: the entropy difference
between the zero field superconducting state and normal state.}
\end{figure}

Figure \ref{fig:HC}(a) shows the specific heat coefficient C/T as a
function of T$^2$ from 0.5 K to 5 K for LaO$_{0.9}$F$_{0.1}$NiAs in
a set of external fields. At zero field, the bulk nature of
superconductivity and the high quality of the sample are confirmed
by the steep jump of C/T at T$_c$=3.8 K, consistent with the
resistivity measurement. A small field of 0.05 T suppresses slightly
the transition temperature, but substantially the specific heat jump
at T$_c$. A small upturn in C/T at low temperature is observed at
0.5 T, which can be attributed to the Schottky anomaly resulting
from the contribution of a small amount of magnetic impurities.

Above T$_c$, C/T shows a good linear T$^2$ dependence. By fitting
the normal state specific heat C with the formula C=$\gamma_n $T +
$\beta $ T$^3$, we find that $\gamma_n $ =4.75 mJ/mol.K$^2$ and
$\beta $ = 0.808 mJ/mol.K$^4$. If these parameters are used to
extract the phonon contributions below T$_c$, we find that the
entropy difference between the superconducting and normal states is
not conserved above T$_c$. This means that the above formula of the
specific heat, i.e. C=$\gamma_n $T + $\beta $ T$^3$, might be too
simple to account for the experimental data below T$_c$ and the
value of $\gamma_n$ above obtained is not correct.

To resolve the above problem, we apply an 10 T magnetic field to
suppress completely the superconductivity as well as the low
temperature Schottky anomaly in the specific heat. Thus the
specific heat at 10 T contains only the contribution from the
normal state electrons and phonon excitations. By subtracting
these normal state contribution at 10 T from the zero-field
specific heat (Fig. (\ref{fig:HC}b)), the normal state electronic
specific heat coefficient is now found to be
$\gamma_n$=7.3mJ/mol.K$^2$. This value of $\gamma_n $ is larger
than that estimated from the normal state data by simply
subtracting a $T^3$ phonon contribution. It is smaller than the
corresponding value for other layered superconductors
NaCoO$_2$($\sim$ 24 mJ/mol K$^2$)\cite{Jin, Luo}, Sr$_2$RuO$_4$
($\sim$ 40 mJ/mol K$^2$)\cite{Nishizaki}, and Li$_x$NbS$_2$
($\sim$ 10 mJ/mol K$^2$,\cite{Dahn}, but larger than that of
Li$_x$NbO$_2$($\sim$ 3.59mJ/mol K$^2$).\cite{liu} In the absence
of magnetic field, the low temperature C/T extrapolates to a small
but finite value of $\gamma_s$=1.16 mJ/mol.K$^2$. This residual
specific heat indicates that there is a residual density of states
at the Fermi level. From this residual specific heat, the
superconducting volume fraction of the sample is estimated to be
about ($\gamma_n-\gamma_s)/\gamma_n$=84$\%$. This value of
superconducting volume fraction is rather high compared with other
previously reported Fe or Ni-based
superconductors.\cite{Kamihara06, Kamihara08, chen, Watanabe}

Figure \ref{fig:HC}(b) shows the difference of the specific heat
coefficient between 0T and 10T, (C$_{0T}$-C$_{10T}$)/T, as a
function of $T/T_c$. The inset of Fig. 4(b) shows the entropy
difference between the superconducting and normal states. The
entropy is now conserved above T$_c$. The normalized specific heat
jump at T$_c$ of LaO$_{0.9}$F$_{0.1}$NiAs is found to be $\Delta C
/(\gamma_n -\gamma_s)T_c$ = 1.9, which is significantly larger than
the value predicted by the weak-coupling BCS theory. This large
specific heat jump may result from strong electron-phonon coupling.
If the the electron-phonon mechanism of superconductivity is
assumed, one can then estimate the electron-phonon coupling constant
$\lambda$ from the modified McMillian formula \cite{allen, carbotte,
cava},
\begin{equation}
\lambda =\frac{1.04+\mu^{\ast}ln(\omega_{ln}/1.2T_c)}{(1-0.62
\mu^{\ast})ln(\omega_{ln}/1.2T_c)-1.04} \label{lambda}
\end{equation}
where $\mu^{\ast}$ is a Coulomb pseudopotential and $\omega_{ln}$ is
a logarithmic averaged phonon frequency. $\omega_{ln}$ can be
determined from the specific heat jump at T$_c$ using the formula:
$\Delta C /\gamma_n T_c =
1.43[1+53(T_c/\omega_{ln})^2ln(\omega_{ln}/3T_c))$. Taking
$\mu^{\ast}$=0.10 and T$_c$=3.8 K, we obtained $\omega_{ln}$ = 61.6
K and $\lambda$=0.93. The large value of $\lambda$ confirms the
strong-coupling nature of the superconducting pairing.

The value of $\lambda$ can be also estimated from the effective mass
normalization. From the first-principles calculations, it was found
that the bare electronic specific heat coefficient for LaONiAs to be
$\gamma_0$= 3.81mJ/mol.K$^2$\cite{fang}. By substituting this value
into the formula $\gamma_n =(1+\lambda) \gamma_0$, we find that
$\lambda$ = 0.92 if LaO$_{0.9}$F$_{0.1}$NiAs to have the same value
of $\gamma_0$ as LaONiAs. This value of $\lambda$ agrees well with
that estimated from the specific heat jump at T$_c$.

The specific data below T$_c$ can be used to determine the pairing
symmetry.\cite{liu} Figure \ref{fig:HC}(b) compares the
experimental data of LaO$_{0.9}$F$_{0.1}$NiAs with the BCS mean
field results for a single-band s- or d-wave superconductors.
Clearly the theoretical curves for both the s- and d-wave
superconductors deviate significantly from the experimental data.
As LaONiAs is a multi-band system, the deviation is most likely to
be due to the multi-gap effect. Moreover, a tiny board shoulder is
observed at T=1.35K (indicated by an arrow in Fig.
\ref{fig:HC}(b)). This is a typical behavior of a multi-band
system, since the superconducting gap may vary at different bands
and the smaller gap can have more contribution to the low
temperature specific heat.

In summary, high quality LaO$_{1-x}$F$_x$NiAs superconducting
samples with extremely narrow superconducting transition width of
$\sim$0.05K and high superconducting volume fraction are
synthesized. The upper critical field is found to be 4.6 T for
LaO$_{0.9}$F$_{0.1}$NiAs. A very steep specific heat jump($\Delta$C)
is observed at transition temperature, and $\Delta C /\gamma_n T_c$
= 1.9 is much larger than the value obtained from the weak-coupling
BCS theory. This large specific heat jump indicates that the
superconducting pairing is in the strong-coupling regime if the
electron-phonon mechanism is assumed. The deviation of the specific
heat data from the BCS result for the single band s- or d-wave
superconductor suggests that this material is a
multi-superconducting gap system.

\begin{acknowledgments}
We would like to acknowledge Z. Fang and L. Lu for the helpful
discussions and H. Chen for the help in the x-ray diffraction
experiments. This work is supported by National Science Foundation
of China, the Knowledge Innovation Project of Chinese Academy of
Sciences, and 973 project of MOST.
\end{acknowledgments}


\begin{thebibliography}{20}


\bibitem{Kamihara06} Y. Kamihara, H. Hiramatsu, M. Hirano, R. Kawamura, H. Yanagi,
T. Kamiya, and H. Hosono, J. Am. Chem. Soc. 128, 10012 (2006).

\bibitem{Watanabe} T. Watanabe, H. Yanagi, T. Kamiya, Y. Kamihara, H.
Hiramatsu, M. Hirano, and H. Hosono, Inorg. Chem. \textbf{46},
7719 (2007)

\bibitem{Kamihara08} Y. Kamihara, T. Watanabe, M. Hirano, and H. Hosono, J. Am. Chem. Soc. xxx, xxxx (2008).

\bibitem{Lebegue}S. Leb\'{e}gue, Phys. Rev. B 75, 035110(2007)

\bibitem{Singh} D. J. Singh and M. H. Du, unpublished, condmat/
arXiv:0803.0429v1.

\bibitem{fang} G. Xu, W. Ming, Y. Yao, X. Dai, S. C. Zhang, and Z.
Fang, cond-mat/arXiv:0803.1282v2.

\bibitem{Haule} K. Haule, J. H. Shim, and G. Kotliar,  unpublished,
cond-mat/arXiv:0803. 1279

\bibitem{Ma} F. J. Ma and Z. Y. Lu, cond-mat/arXiv:08033.2358.

\bibitem{chen} G. F. Chen, Z. Li, G. Li, J. Zhou, D. Wu, J. Dong, W.
Z. Hu, P. Zheng, Z. J. Chen, J. L. Luo, and N. L. Wang ,
unpublished, preprint: cond-mat/arXiv:0803.0128v1.

\bibitem{wen}G. Mu, X. Zhu, L. Fang, L. Shan, C. Ren, H.H. Wen
unpublished, cond-mat/arXiv:0803.0928v1;  X. Zhu, H. Yang, L. Fang,
G. Mu, H.H. Wen, unpublished, cond-mat/arXiv:0803.1288.

\bibitem{Dong} J. dong et al., unpublished, cond-mat/arXiv:0803.3426.

\bibitem{Dahn} D. C. Dahn, J. F. Carolan, and R. R. Haering, {Phys. Rev. B} {\bf 33}, 5214 (1986).

\bibitem{Jin} R. Jin, B. C. Sales, P. Khalifah, and D. Mandrus, {Phys. Rev. Lett.} {\bf 91}, 217001 (2003).

\bibitem{Luo} J. L. Luo, N. L. Wang, G. T. Liu, D. Wu, X. N. Jing, F. Hu, and T. Xiang, {Phys. Rev. Lett.} {\bf 93}, 187203 (2004).

\bibitem{Nishizaki} S. Nishizaki, Y. Maeno, S. Farner, S. Ikeda, and T. Fujita, {J. Phys. Soc. Jpn.} {\bf 67}, 560 (1998).

\bibitem{liu} G. T. liu, J. L. Luo, Z. Li, Y. Q. Guo, N. L. Wang, and D.
Jin, Phys. Rev. B, \textbf{74}, 01250492006).

\bibitem{allen} P. B. Allen and R. C. Dynes, Phys. Rev. B \textbf{12}, 905(1975).

\bibitem{carbotte} J. P. Carbotte, Rev. Mod. Phys \textbf{62}, 1027(1990).

\bibitem{cava} T. Klimczuk, F. Ronning, V. Sidorov, R. J. Cava and J. D. Thompson, Phys. Rev. Lett \textbf{99}, 905(2007).

\end{thebibliography}
\end{document}